\documentstyle[preprint,aps,eqsecnum]{revtex}
\tightenlines
\begin{document}
\draft
\title{ Coulomb effects on the transport properties of 
quantum dots in strong magnetic field}
\vspace {1cm}  
\author{V.\ Moldoveanu, A.\ Aldea, A.\ Manolescu, M.\ Nit\~{a}} 
\address{National Institute of Materials Physics, POBox MG7,
Bucharest-Magurele, Romania}

\maketitle 
\par
\begin{abstract}

We investigate the transport properties of quantum dots placed in strong
magnetic field using a quantum-mechanical approach based on the
2D tight-binding  Hamiltonian with direct Coulomb interaction and the
Landauer-B\"{u}ttiker (LB) formalism. The electronic transmittance and the 
Hall resistance show Coulomb oscillations and also prove multiple addition
processes. We identify this feature as the 'bunching' of electrons observed
in recent experiments and give an elementary explanation in terms of
spectral characteristics of the dot. The spatial distribution of the 
added electrons may distinguish between edge and bulk states and it has 
specific features for bunched electrons.
The dependence of the charging energy on the number of electrons is discussed both
for strong and vanishing magnetic field. The crossover from the tunneling
to quantum Hall regime is analyzed in terms of dot-lead coupling.

\end{abstract}
\vspace*{-0.25cm}

\section{Introduction}
\par
The basic phenomena in quantum dots (QD) are usually described by the 'orthodox 
theory' \cite{K,B} which explains the Coulomb blockade effects (charge 
quantization and the oscillations of the electrical resistance) in terms of the capacitive 
properties of the isolated (or very weakly coupled) dot. Generally speaking, 
the corresponding capacitance should depend on the dimension and the dot shape, 
number of electrons accommodated inside and the electron-electron interaction 
(EEI). Nevertheless, the orthodox theory considers the dot capacitance as 
being a constant, independent of the number of electrons and the size 
quantization effects. 
This idea can be accepted for large metallic QDs when the charging energy 
due to the Coulomb interaction is larger than the level spacing of the 
one-electron energy spectrum. However many specific properties of QDs 
remain beyond this model. Such aspects occur especially for small semiconductor 
quantum dots, when the level spacing is relatively large, the distribution 
of the energy levels depends visibly on the dot shape and the number 
of electrons is smaller then in the metallic case. In small dots the 
interplay between the quantum aspects and the charging effects is important. 
This is why a more advanced description pretends to pay attention to the 
one-particle energy spectrum and to consider a more realistic Hamiltonian.
\par 
In spite of the general acceptance that the charging energy should depend on 
the number of electrons inside dot $N$, fact which is proved experimentally by
the irregular Coulomb blockade oscillations \cite{T}, this effect has been 
simulated numerically only recently \cite{Macucci,LOC99}.  
The situation becomes even more complicated when the coupling to the leads 
is taken into account carefully, eventually in the presence of a magnetic field.
In fact, this coupling between dot and leads (or using other words: the degree
of pinching/constrinction at the contacts) decides the degree of quantization 
of the charge in QD and affects the electronic transmittance through the 
dot and the Hall resistance. Generally, in the literature, the lead-dot (LD) 
coupling is taken into onsideration assuming an electronic transition rate
$\Gamma$ between leads and dot.
 This phenomenological way of thinking is replaced here by the explicit 
calculation of the transmittance properties in the quantum-mechanical 
Landauer-B\"uttiker formalism. This formalism works as long as the EEI is 
present in the dot but is neglected in the leads \cite{MW}.
In this framework we show that the LD coupling plays an even more pregnant 
role in the presence of a strong magnetic field, in which case an interesting 
crossover from Coulomb oscillations to the Quantum Hall regime can be noticed 
for the transverse resistance with increasing coupling. 
The strong magnetic field perpendicular on the two-dimensional dot gives rise 
to edge-states even if the dot is small. Then, for large LD coupling, QHE-type 
effects appear, although, usually, only the first plateau is 
visible in small dots (and not always very clean). Such aspects were evidentiated 
for the first time in \cite{Hart} by the use of a tight-binding(TB) model in 
the absence of the electron-electron interaction.
\par
The philosophy of our approach for calculating the transport properties of 
quantum dots is the following: the QD is coupled weakly to four semi-infinite 
leads supporting many channels. The number of degree of freedom of the 
terminals is infinite while the QD has only a finite number of degree of 
freedom, so that the Fermi level $E_F$ of the whole system is imposed by 
terminals. At a given magnetic field, when a gate potential $V_{g}$ is applied 
and varied, the fixed $E_F$  scans the whole energy spectrum.
Without magnetic field, aspects concerning Coulomb effects and the 
importance of size quantization have been discussed in \cite{LOC99} in the 
Hartree approximation: the calculated addition spectrum shows   
the dependence of the charging energy on the number of particles,
the electronic transmittance peaks --- which  correspond to the 
charge-degeneracy points --- are distributed irregularly keeping track of 
the size quantization. The width of the transmittance peaks depends not only 
on the LD coupling but also on the strength of the EEI, showing the 
cotunneling effect at resonances.
\par
The aim of this paper is to discuss  the transport
properties of the open dots in the regime of a strong magnetic field. 
We calculate the addition spectrum, the transmittance matrix and 
the Hall resistance. We show that, when the contacts are pinched, the quantum 
plateaus disappear and quantum oscillations of $R_H$ are installed even in the 
absence of EEI. Next we show the characteristics of the $R_H$ oscillations in 
the presence of the long-range direct electron-electron interaction in 
Hartree approximation. These oscillations as function of $V_g$ or magnetic 
flux were observed experimentally already long time ago \cite{Ford}, but 
they were never simulated numerically on the basis of a theoretical model. 
Our calculations put into evidence the electronic {\it bunching} in the  
addition process, an effect which was found recently by single electron 
capacitance spectroscopy(SECS) \cite{bunching}.
\par
The description of the formalism is made in the next section, while in 
Sec. III we discuss briefly  some  aspects of resonant transport through  a
non-interacting QD subject to a strong magnetic field. Our main results are 
established and commented in Sec. IV, the conclusions being isolated in Sec. V.
\par 
\vspace*{-0.25cm}
\section{The formalism}
We use a pure  quantum-mechanical approach of the transport properties of open 
quantum dots, which is based on the Landauer-B\"{u}ttiker formalism and the 
Hartree approximation for the electron-electron interaction. While the Hartree 
term is meaningless in infinite homogeneous system it becomes important for 
finite systems like QD.
The method is complementary to the semi-classical master-equation approach 
\cite{B}, goes beyond the constant-interaction model and is  able to 
account for size, tunneling and interaction effects in quantum dots in 
the presence of the magnetic field.
\par
Our calculations are based on a lattice model. In spite of the 
fact that in recent years this model is extensively used for the study of 
various effects in quantum dots\cite {10,Canali} we would like to say a few 
words of caution on this approach to small systems, in the presence of the EEI.
When the lattice model is considered as the discretization of a continuous 
system by with a rectangular grid of intersite distance $a$, the (direct) 
EEI reads $U\sum_{j>i}{1/ |i-j|}c{\dagger}_ic{\dagger}_jc_jc_i$~,
where $i,j$ are the lattice sites and $U=e^2/a$; this means that the strength 
parameter $U$ depends on the grid which is meaningless. Also the hopping 
integral  $t=\hbar^2/2ma^2$ and the radius $r_s=e^2/(at\sqrt{4\pi\nu}$ ($\nu$ 
being the filling factor) depend on the grid. 
One may say that the discretization works better at low filling factor 
(meaning  small $a$ or large number of grid sites).
Another way to avoid this paradox is to consider the lattice model as a 
tight-binding approximation, i.e. to assume that the overlap between the 
atomic orbitals (and usually, one considers only one type of orbitals) 
located on different sites is small and the effective mass is big. 
In general, this is not the case for the semiconductors used in 
the experimental devices. In particular, the strength of the 
electron-electron Coulomb interaction depends on the dielectric constant 
of the semiconductor host material, which we incorporate here in 
our coupling constant U.
\par
The discrete model allows the tailoring of different shapes and introducing 
the magnetic field as a phase of the hopping integral. The parameters 
controlling the problem are:  
a) strength of the $LD$ coupling, 
b) size and shape of the dot and the magnetic flux, which all of them, 
determine the electronic spectrum in the absence of the EEI, c) strength
of the EEI ($U$). 
\par
We model the QD as a 2D mesoscopic plaquette weakly coupled to four external 
semi-infinite leads. In the TB approximation the Hamiltonian is written as: 
\begin{equation}
H=H^{D}+\sum_{\alpha}H_{\alpha}^{L} + \sum_{\alpha}H_{\alpha}^{LD} 
\end{equation}
In the above relation $H^D$ is used to describe the isolated QD, 
$H^L_{\alpha}$ characterize the lead $\alpha$ ($\alpha$ =1,..,4), while the 
lead $\alpha$ is coupled to the dot by:
\begin{equation}
H_{\alpha}^{LD}=t^{LD}(c_{0\alpha}^{\dagger}c_{\alpha}+
c^{\dagger}_{\alpha}c_{0\alpha}) ,
\end{equation}
where the operator $c^{\dagger}_{\alpha}$ creates an electron in the dot state 
$\vert\alpha\rangle$ and $c_{0\alpha}$ annihilates it in the neighboring lead 
state  $\vert0\alpha\rangle$.
Here $t^{LD}$ is the hopping integral between dot and leads.
Since the role of the leads is only to inject and drain the electrons or to 
probe the potential drop, the EEI will be included only in the Hamiltonian of 
the dot. 
In the 'orthodox theory' the Coulomb effects are mostly studied in the 
constant-interaction model (see, for instance, \cite{Aleiner}) which considers 
the Hamiltonian $H_{ee}=(e^2/2C)(N-N_g)^2$ ($C$ = capacitance, $N$ = operator 
of the total number of particles, $N_g$ = external parameter related to the 
gate potential). Here, we use a long range direct Coulomb interaction. 
Expressed in terms of creation and annihilation operators on localized states 
indexed by $i\in QD$, the Hamiltonian of the dot reads: 
\begin{equation}
H^D=\sum_{i,j}(t^D_{ij}~c^{\dagger}_{i}c_{j}+
\frac{1}{2}U_{ij}c_i^{\dagger}c_j^{\dagger}c_jc_i)
+\sum_{i}V_g c_i^{\dagger}c_i~,~~i,j\in QD .
\end{equation}
The  external gate $V_g$ is simulated by a site energy in $H^{D}$.
In Hartree approximation and  nearest-neighbors model, it becomes:
\begin{equation}
H^{D}=\sum_{i} 
\Big(V_{g}+U\sum_{j>i}{<n_{j}>\over |i-j|}\Big)
c_i^{\dagger}c_{i}
+ t^{D}\sum_{<i,j>}e^{i2\pi\phi_{ij}}~c_i^{\dagger}c_{j}.
\label{hdot}
\end{equation}
where $<n_{j}>=<c_j^{\dagger}c_j >$ is the mean occupation number of the 
site $j$ and $U$ is the parameter describing the strength of the EEI.
We have chosen $t^{D}=1$, i.e. the energy unit is the hopping integral in QD, 
and we have denoted by $<...>$ the nearest-neighbors summation.  
The Peierls phase $\phi_{ij}$ is proportional to the magnetic flux through the
unit cell $\phi=Ba^2$ measured in quantum flux units $\phi_0$ (the explicit 
expression of $\phi_{ij}$ depends on the site indexes i,j and on the gauge we 
choose \cite{AA}).
\par
At this point a useful "trick" is to describe the open dot by an effective 
Hamiltonian which includes the influence of the leads. 
Eliminating formally the degrees of freedom of the leads, one obtains
a non-Hermitean Hamiltonian depending on the energy:
\begin{equation}
H_{eff}^{D}(z)= H^{D}+H^{DL}{1\over z-H^{L}}H^{LD}= 
H^{D}+H^{DL}G^L(z)H^{LD} .
\end{equation}
\par
In the above equation, the Green function $G_L(z)$ of the semi-infinite lead 
can be calculated analytically:
\begin{equation}
G^L_{ij}(z)=\frac{1}{t_L(\zeta_2-\zeta_1)}\cdot[\zeta_1^{|i-j|}-\zeta_1^{i+j+2}] 
~,~~~~~i,j\in lead
\end{equation}
\par
where $\zeta_1$ and $\zeta_2$ are the roots of the equation:
\begin{equation}
t_L\zeta^2-z\zeta+t_L=0~, \hspace*{1cm} |\zeta_1|<1<|\zeta_2| \,.
\end{equation}
\par
Let $\zeta_1(z)$ in the upper half plane. By approaching the real axis from 
above one obtains $\zeta_1(z)=\zeta(E+i0)=e^{-ik}$, where $k$ is defined by 
$2t_L\cos k=E$, $t_L$ being the hopping energy of leads. After straightforward
manipulations the effective Hamiltonian of the dot is obtained explicitly:
\begin{equation}
H_{eff}^D=H^{D}+\tau^2t_L\sum_{\alpha}~e^{-ik}c^{\dagger}_{\alpha}c_{\alpha}
\end{equation}
The ratio $\tau=t_{LD}/t_L$ defines the degree of constrinction at the contacts
and represents an input parameter which can be varied  continuously. 
It is important to observe that the influence of the leads is expressed as a 
non-Hermitean diagonal term proportional to $\tau^2$, which produces a shift 
in the real part of the eigenvalues of $H^D$ and introduces also an imaginary 
part. If $\tau\ll 1$, i.\ e. for a weakly-coupled dot, these shifts become 
negligible and the spectrum of $H_{eff}$ approaches the spectrum of the 
isolated dot. This behavior has important consequences on the conductivity 
matrix $g_{\alpha\beta}$, seen as the transmittance $T_{\alpha\beta}$,  
which can be expressed in terms of the retarded Green function 
$G^+(E)=(E-H_{\rm eff}+i0)^{-1}$ by the LB formula:
\begin{equation}
g_{\alpha \beta}~=~{e^2\over h} T_{\alpha\beta}~
=4\frac{e^2}{h}~\tau^4t_L^2{\sin^2}k~ |G_{\alpha\beta}^+(E_F)|^2 \,,
~~\alpha\not=\beta \,.
\end{equation}
Once the conductance matrix $g_{\alpha\beta}$ is known, the Hall resistance can
be calculated immediately,\cite{Bu}
\begin{equation}
R_H=(g_{21}g_{43} - g_{12}g_{34})/D ~,
\end{equation}
where $D$ is a $3 \times 3$ subdeterminant of the $4 \times 4$ 
matrix $g_{\alpha\beta}$. The matrix elements of the Green's function,
$G^{+}_{ij}(E)$, are calculated numerically using the self-consistency condition:
\begin{equation}
<n_{j}>={1\over\pi}\int_{-\infty}^{E_{F}} {\mathrm Im}~G_{jj}^{+}(E)\,dE\,.
\end{equation}
\par
In the weak-coupling limit $\tau\ll 1$, the transport problem 
reduces to a tunneling problem. Indeed, from Eq.(9) it follows that the 
poles of the Green function will induce a series of peaks in the 
transmittance, and Eq.(11) shows that the mean number of electrons in the QD
changes abruptly by 1 at every peak, indicating a charge addition process. 
So, the correspondence between the peaks observed in the transmittance and 
the charge accumulation in the dot is manifestly established. 
\par
Taking into account the strong conditioning of the transport properties 
(electronic transmittance and Hall resistance) by the energy spectrum of the
dot, we have to perform  a comparative analysis of spectral properties
of $H^D$, with and without EEI.

\section{Non-interacting dot in strong magnetic field}
In this section we address the resonant transport through a
non-interacting QD, mainly because this simple framework gives a clear
picture of the constrinction effects. Moreover some data about the 
non-interacting spectrum will be needed in the next section.           
\par
As it is known, when periodic boundary conditions are imposed to a 
{\it non-interacting} 2D electronic system subjected to a perpendicular 
magnetic field, the tight-binding approach yields the Harper equation 
associated with the usual Hofstadter-butterfly spectrum.
When the periodic boundary conditions are replaced by the Dirichlet
conditions, a 'quasi-Hofstadter' spectrum is obtained \cite{AA}, 
since the hard-wall potential lifts the degeneracy and the gaps get 
filled with eigenvalues which correspond to the so-called edge states 
(which are extended along the edges of the system and 
are responsible for the quantization of the Hall conductance ).
Another type of states are the 'bulk states' which are grouped in energy bands 
and, geometrically, concentrated in the middle of the dot. The nature - bulk 
or edge- of a given state $\Psi_{n}(\phi)$ can be checked also by its 
chirality \cite{AAN}, i.e. by the sign of the current carried by that state, 
defined as the slope of the energy level:
\begin{equation}
I_n=\frac{d E_n(\phi)}{d \phi}.
\end{equation}
Since the eigenvalues $E_n(\phi)$ are not monotonic functions of the magnetic 
flux (as it can be noticed in Fig.1a), it follows that the nature of 
the corresponding eigenstate may change from bulk to edge or vice-versa
when the flux is varied.
\par
For strongly pinched contacts, a continuous variation of the 
gate potential (or, equivalently, of the Fermi level) gives rise to  
a resonance peak whenever the Fermi level is aligned to an eigenvalue of 
the isolated dot (the width of the peak is determined only by the strength of 
the LD coupling in the noninteracting case). 
This can be seen in Fig.1b which depicts the transmittance 
spectrum of the non-interacting quantum dot as function of $E_F$ mapped onto 
the corresponding piece of the quasi-Hofstadter spectrum. 
\par
The modifications in the transmittance induced by pinching is shown in Fig.2 
for $T_{12},T_{13}$ and $T_{14}$, in the case of strong magnetic field.
One remarks that for completely open QDs (at $\tau=1.0$) the transmittances 
take the values which describe  the Quantum Hall regime: all $T_{\alpha,\beta}$ 
with $\alpha \ne \beta$ vanish except $T_{\alpha,\alpha+1}$ \cite{Bu}.
On the other hand, for very weakly coupled QDs ($\tau\ll 1$), which corresponds
to the resonant tunneling regime, the dwell time of the electron inside the 
dot increases and all $T_{\alpha,\beta}$ become of the same order of magnitude.  

\par
While the transmission spectrum identifies the positions of the levels it 
cannot specify whether the corresponding states are edge- or bulk-type. This 
can however be evidentiated by the Hall resistance for simple reasons: if the 
leads are strongly coupled to the dot, the Hall resistance exhibits quantum 
Hall plateaus in the range of the spectrum occupied by  edge states. At strong 
constrinction, interference effects occur when the electron travel along the 
edge states, resulting in oscillations of the resistance in the region of the 
former plateau. This is shown in Fig.3a where each minimum in the Hall 
resistance corresponds to a resonance condition (when the Fermi level equals 
an eigen-energy belonging to an edge state). Remark the sudden drop of the 
Hall resistance between the QH plateaus indicating a narrow bulk domain. 
\par
When electron-electron interaction is considered, some novel features appear 
that can be traced from Fig.3b, which depicts $R_H$ versus $V_g$ for $U=0.5$. 
The discussion of this figure is postponed  to the next section where the 
influence of the EEI on the edge and bulk states will be analyzed.


\section{The interacting case.}
When the electron-electron interaction is taken into account, important 
differences appear in the positions and widths of the transmittance 
peaks and simultaneously in the Hall resistance. This is due to the Coulomb
blockade effect, meaning that the addition of an extra electron needs some 
energy which is not simply the difference between two consecutive one-
electron  levels,
but has also a contribution coming from the electron repulsion. 
We shall start with some considerations on the addition spectrum which 
will be usefull for understanding the features of the resonant transport 
through QDs in the presence of the interaction .
Let $E_n(N)$ be the n-th eigenvalue of the system containing $N$ electrons.
$E_n(N)$ has a monotonic dependence on the Fermi energy.
When $E_F$ (the diagonal line in Fig.4) approaches $E_n(N)$ the addition of 
the $N+1$-th electron becomes possible and the whole spectrum raises 
with the {\it charging energy}:
$E_{ch}(N,N+1)= E_n(N+1)$-$E_n(N)$.
One notices from Fig.4 that the charging energy is not supplied 
step-like but linearly (with slope=1.0) along an interval $\delta E_F=E_{ch}$ ; the addition of the 
extra electron occurs at the {\it charge degeneracy point} situated in the 
middle of this interval, where $E_F=(E_{n}(N+1)-E_{n}(N))/2$.
The transmittance shows the addition spectrum properties: the peaks point 
versus the degeneracy points; their widths -measured at the bottom- 
equals the charging energy and are due to the so-called 'co-tunneling' near 
the degeneracy points \cite{Sch}).


\par
Now we turn to discuss Fig.3b and to make the comparison with the Fig3.a 
(interacting vs. non-interacting case). The similarities of the two figures 
suggest that the edge states are present also in the interacting case, giving 
rise to  oscillations of $R_H$ on different quantum Hall plateaus. There are 
however qualitative differences: for $U\neq 0$, both the edge and bulk regions 
are much expanded, so that the whole picture is pushed upwards on the energy 
scale (in the numerical calculation this is equivalent to  large negative $V_g$). 
This means that the Coulomb interaction increases the level spacing and a 
striking consequence is the slower drop of $R_H$ in the region of the bulk 
states. However, in order to fully establish the nature of the states we have 
to observe the changing of the local electronic distribution $n_i$, when 
exactly one more electron is added, i.e to calculate:
\begin{equation}
\Delta n_{i}(N, N+1)= n_{i}(N+1)-n_{i}(N) ~,~~i\in QD,~N =integer.
\end{equation}
In order to make sure that $N$ is an integer, one has to calculate Eq.(13)
for those values of $V_g$ which ensures an integer number of electrons in QD ; 
they correspond to two consecutive valleys in the transmittance spectrum Fig.4.
The interest in Eq.(12) follows from the fact that the map of $\Delta n_{i}$  
shows how the $(N+1)-th$ electron is added, namely on the edge or in the bulk. 
For instance, Fig.5a gives a clear proof that the 8-th electron is trapped by 
an edge state. On the other hand, the bulk states are much damaged: this can 
be seen in  Fig.5b where the added electron is more or less distributed 
everywhere. 
Eq.(13) contains the aprioric assumption that the electrons are added 
individually. However, recent experimental results suggest that the electrons 
may enter the dot not only one by one but also in bunches, fact which can be 
seen in the addition spectrum \cite{bunching}. This very 'non-orthodox'
feature has to be noticed also in the transmittance properties of the dot.
So, let us discuss the double peaks existing in the calculated transmittance
in Fig.6a. They become possible when two degeneracy points are very 
close and the co-tunneling effect does not permit their resolution.
The origin of this effect consists in two close poles of the
resolvent $(E-H_{\rm eff})^{-1}$ fact that yields a multiple addition process,
which is nothing else but the 'bunching'. While the bunching is 
not allowed in the range of edge states (which are well-separated), it appears 
in the bulk region due to the existence of quasi-degenerate states. The way in 
which the two grouped electrons are distributed in the dot is shown by 
$\Delta n_{i}(9,9+2)$ in Fig.5c.
We stress that, in this case, the distribution has evident maxima at the corners. 
This corroborates with the results obtained recently by Canali \cite{Canali}. 
The bunching has important consequences on the oscillation amplitude of 
$R_{H}$, in the sense that the amplitude is suppressed whenever a 
bunching appears (see Fig.6b). The bunching is not conditioned by the presence 
of the magnetic field but on the existence of the co-tunneling effect,
what reveals its Coulombian nature.


\par
When the transmittance is plotted against the number of electrons (Fig.7)
interesting features can be noticed : some maxima are reached at half integer 
number of electrons, in accordance with the constant capacitance model
in which the additions becomes costless at the charge degeneracy points
satisfying the relationship $CV_g/e=~N+1/2$ \cite{Glazman}. This is
clearly violated, however, any time the multiple addition process occurs
(see the bunching of 6-th and 7-th electrons in the same figure).
\par
Another interesting effect in the presence of the magnetic field is the 
dependence of the  charging energy on the number of particles shown in Fig.8.
Without field the charging energy depends irregularly on $N$, confirming the 
results obtained in Ref.4. The behavior changes drastically in strong magnetic 
field, showing a monotonic increase of the charging energy as long as the 
electrons are added on the edges (this occurs for $3 \leq N \leq 8$, which 
also correspond to the quantum oscillations of the Hall resistance on the 
first plateau).


\section{Conclusions}
We have studied the transport properties of interacting quantum dots pierced
by a strong magnetic field. The transmittance and the Hall resistance were 
calculated in the Landauer-B\"{u}ttiker formalism, in the tight-binding 
picture (which contains explicitly the dot-lead coupling ), while the 
electron - electron interaction was considered in the Hartree approximation. 
This approach is able to describe the competition between the size, 
interaction and tunneling mechanisms in QD.  
After proving the essential role of the dot-lead coupling we have obtained
the Coulomb oscillations of the transmittance and Hall resistance
for various degrees of constrinction.
The charge degeneracy points were shown to coincide with the minima of
Hall resistance and the peaks of the transmittance.
 It turns out that the edge states are stable against
interaction, only the interlevel spacing being increased. On the contrary
the bulk states are much more damaged. The labelling of states into 
edge- or bulk-like was done by an explicit mapping of the spatial 
distribution for each added electron.
We have also presented an elementary explanation for the bunching of electrons 
recently revealed by SECS experiments. In what concerns the dependence of the
charging energy on the number of electrons, it is shown that in zero
flux case, $E_{ch}$ depends irregularly on $N$ but increases monotonically
in strong magnetic field as long as the Fermi level lies in the region of
edge states.

\acknowledgements

\vspace*{0.25cm} \baselineskip=10pt{\small \noindent  A. M. and A. A.
benefited of  the hospitality and support of the International Centre for  
Theoretical Physics, Trieste, Italy, at the Research Workshop on Mesoscopic 
Systems, and through the Associateship Scheme (A. M.).
A. A. is very grateful to Prof. Johannes Zittartz for his hospitality  at the
Institute of Theoretical Physics, University of Cologne, where part of
this work was performed under SFB-341.  Valuable discussions with 
Dr. P. Gartner are acknowledged.

\begin{figure}
\caption{The correspondence between the transmittance spectrum of a non-
interacting  QD in strong magnetic field ($\phi=0.15$) and the Hofstadter
spectrum. Each transmittance peak arises when $E_F$ equals an eigenvalue
of the isolated spectrum.}
\label{fig1}
\end{figure}

\begin{figure}
\caption{The dependence of transmittances on constrinction ($T_{12}$-full
line, $T_{13}$-dashed line and $T_{14}$-dotted line, $\phi=0.15$).
Pinching effect is obvious: at large coupling the conditions for QHE
are fulfilled, while by continously decreasing LD coupling, $T_{ij}$
evolves smoothly to the same order of magnitude.}
\label{fig2}
\end{figure} 

\begin{figure}
\caption{ (a) Quantum oscillations of the Hall resistance for pinched
contacts ($\tau=0.5$,  $\phi=0.15$, $U=0.0$). Remark the sudden drop of
$R_H$ between different Hall plateaus. (b) Interaction effects on the
Hall resistance ($U=0.5$): the drop of $R_H$ is slower and the oscillations 
in the range of edge states are widened but their amplitude is poorely
affected by EEI.}
\label{fig3}
\end{figure} 

\begin{figure}
\caption{ (a) The Hartree spectrum of an isolated dot in strong magnetic
field ($\phi=0.15$, $U=0.5$). (b) The transmittance of a weakly-coupled
QD ($\tau=0.1$) as a function of the gate potential. The charging energy can be
obtained as the width at the bottom of the peaks.}
\label{fig4}
\end{figure} 

\begin{figure}
\caption{ The spatial distribution of the added electron inside the
interacting dot: (a)$\Delta n_i(7,8)$ - the 8-th electron is added
strictly on edge. (b)$\Delta n_i(8,9)$ - the 9-th electron is distributed 
almost uniformly. (c)$\Delta n_i(9,9+2)$ - a multiple addition process in 
which the 10-th and 11-th are added together in the dot; a clear 
addition is made in the bulk, but some maxima are reached also at 
the corners.}    
\label{fig5}
\end{figure}

\begin{figure}
\caption{ (a) The transmittance spectrum in strong 
magnetic field ($\phi=0.15$, $U=0.5$). Note the appeareance of the double 
peaks associated with multiple addition processes. 
(b) Oscillations of the Hall resistance induced by strong pinching 
($\tau=0.1$). Note that whenever a multiple addition process is allowed the 
oscillation amplitude is nearly vanished.}  
\label{fig6}
\end{figure}

\begin{figure}
\caption{ The dependence of transmittance on number of electrons. Most of the 
the maximas are reached at half-integer numbers of electrons, but this condition 
is not obeyed when the bunching appears.}
\label{fig7}
\end{figure} 

\begin{figure}
\caption{ The charging energy vs. the number of electrons. Without magnetic
field $E_{ch}$ depends irregurarly on $N$, while in a strong magnetic field
 a monotonic increase is observed as long as the electrons are added on 
the edge.}
\label{fig8}
\end{figure}

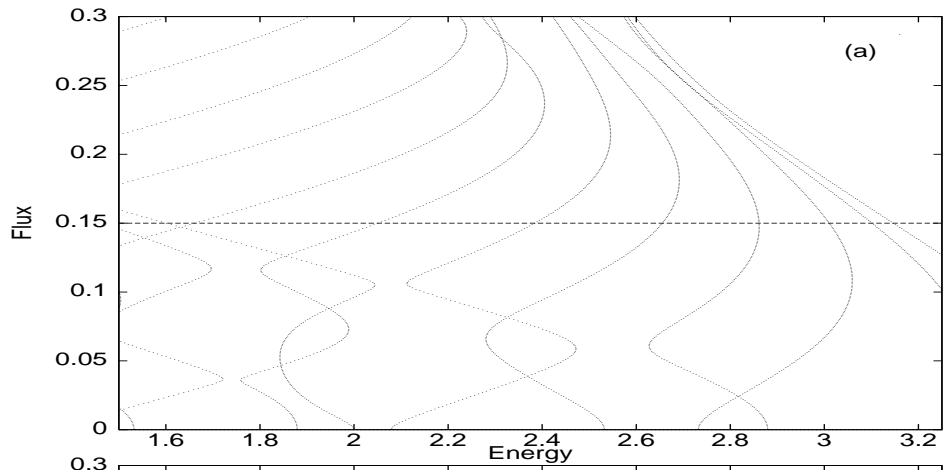
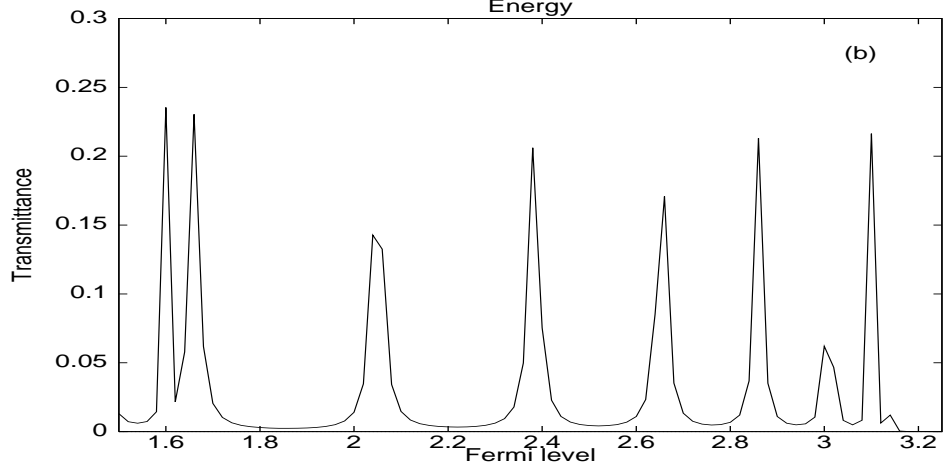
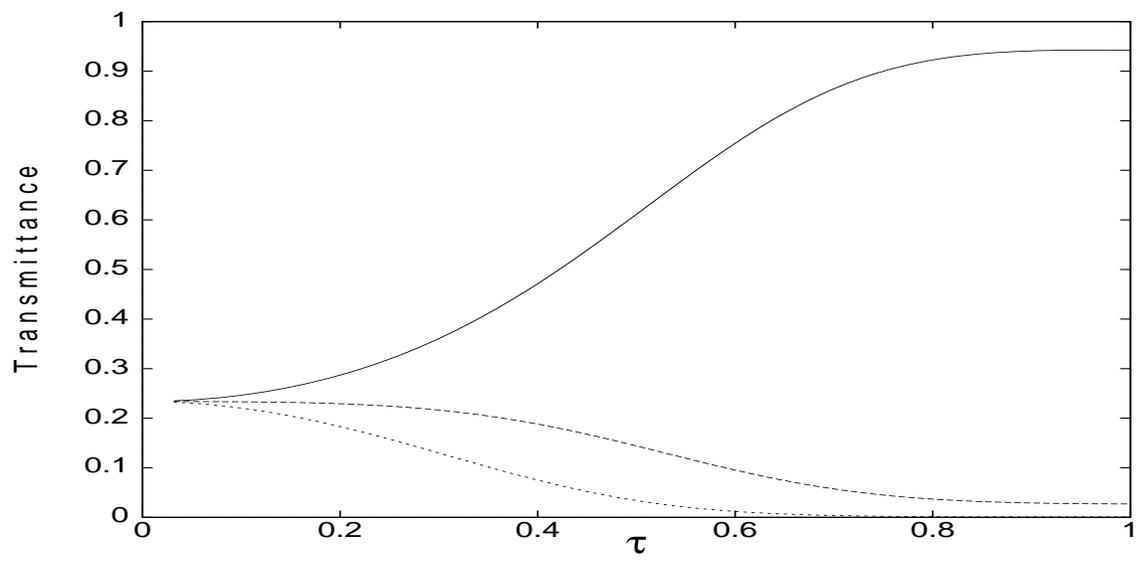
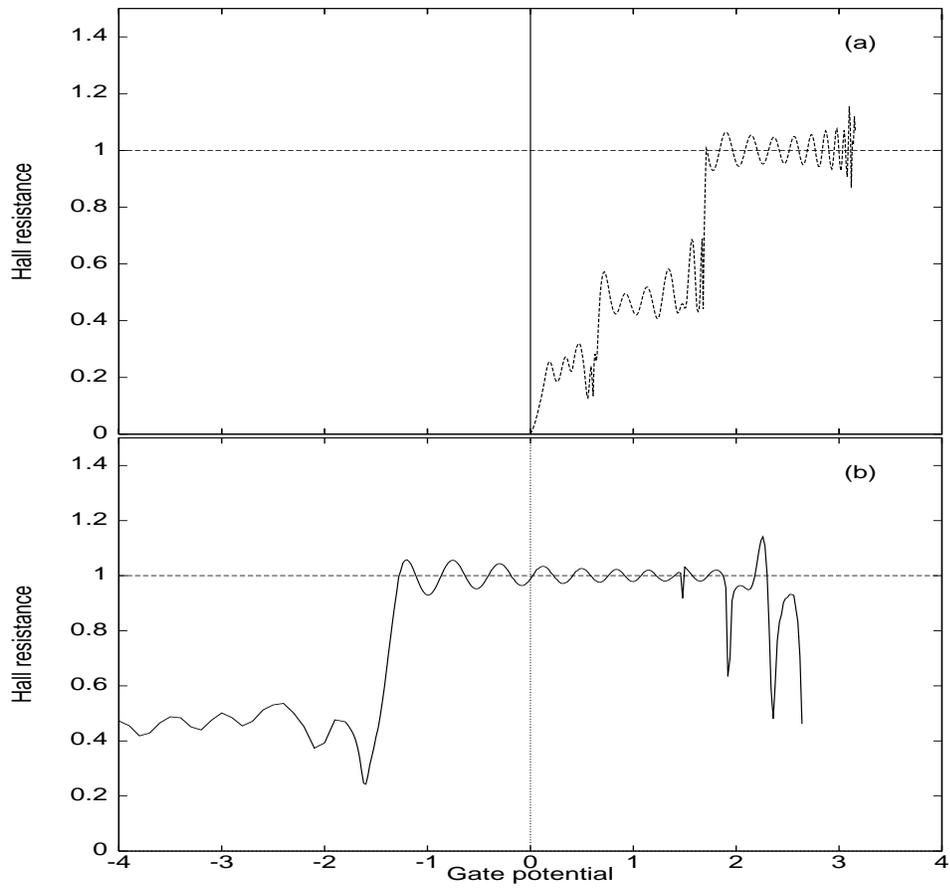
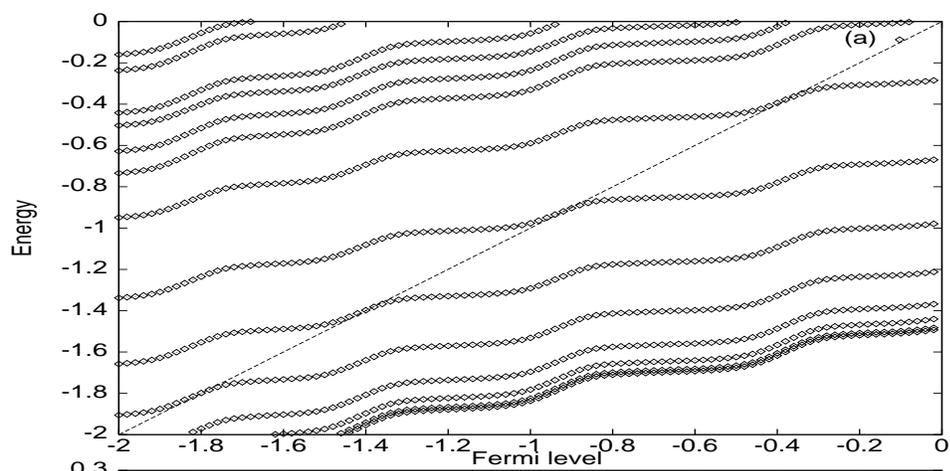
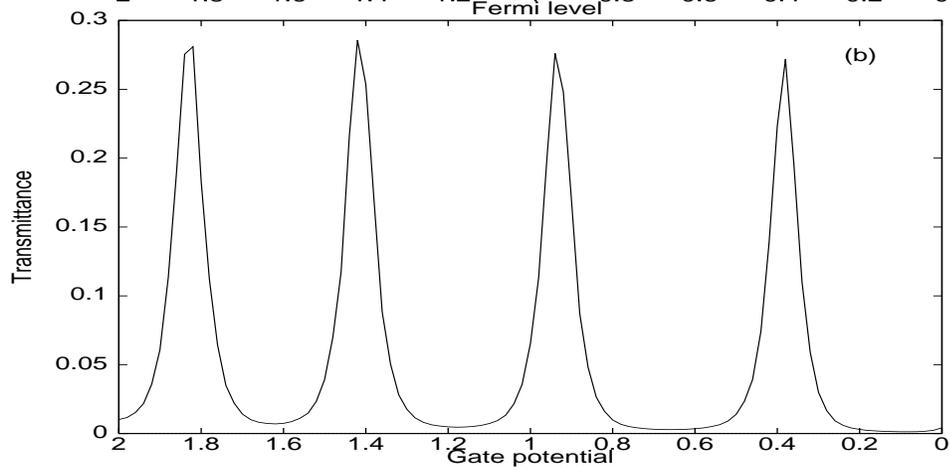
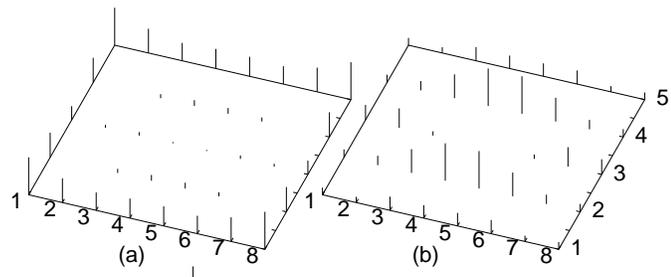
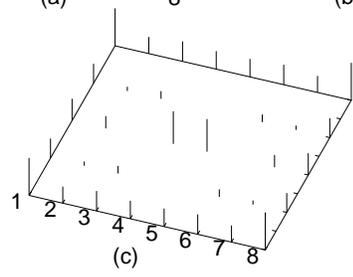
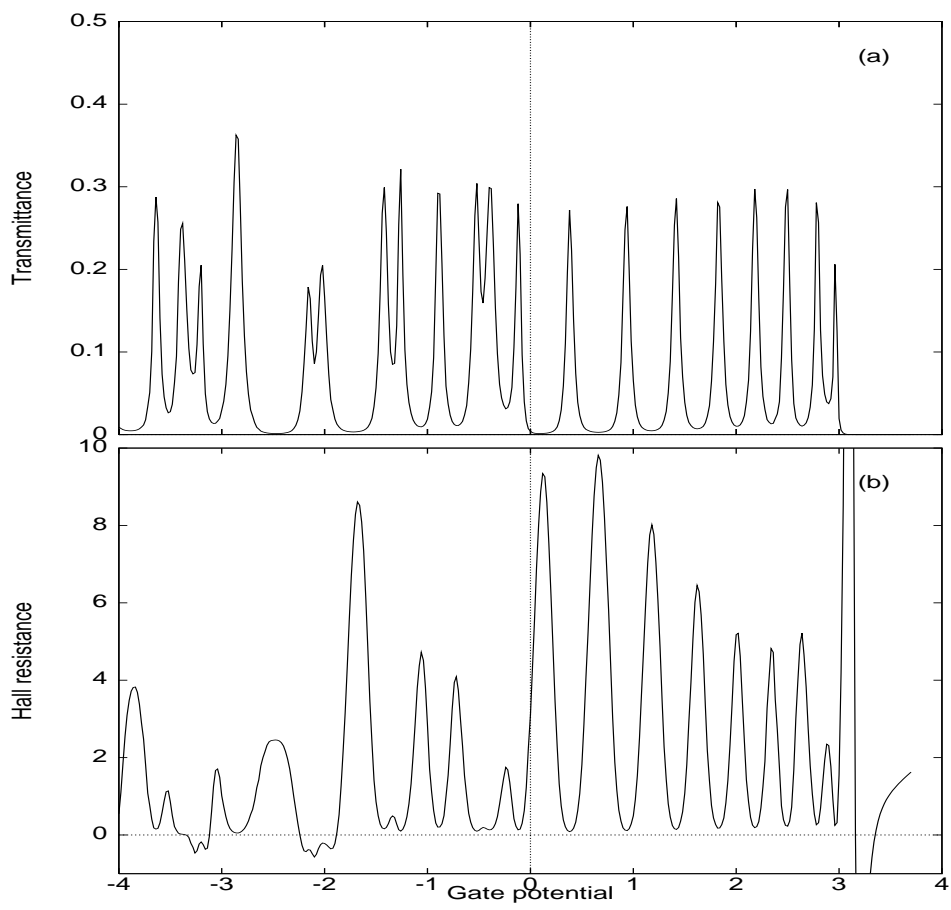
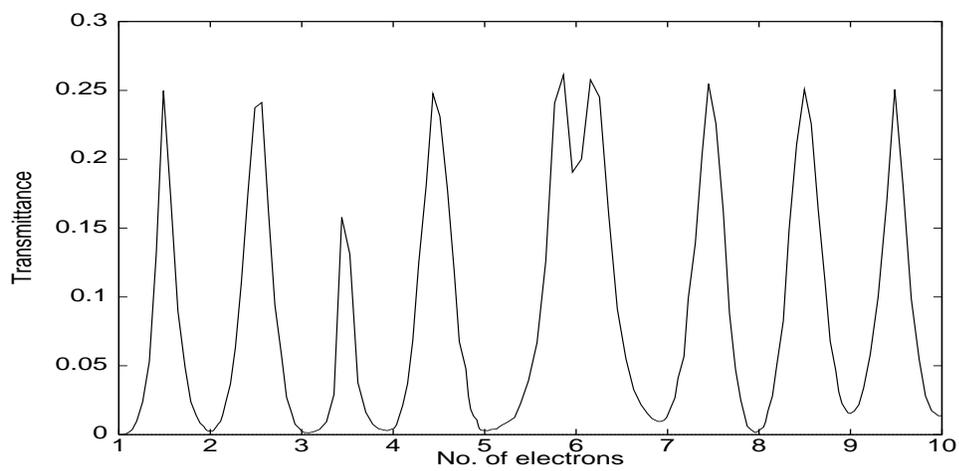
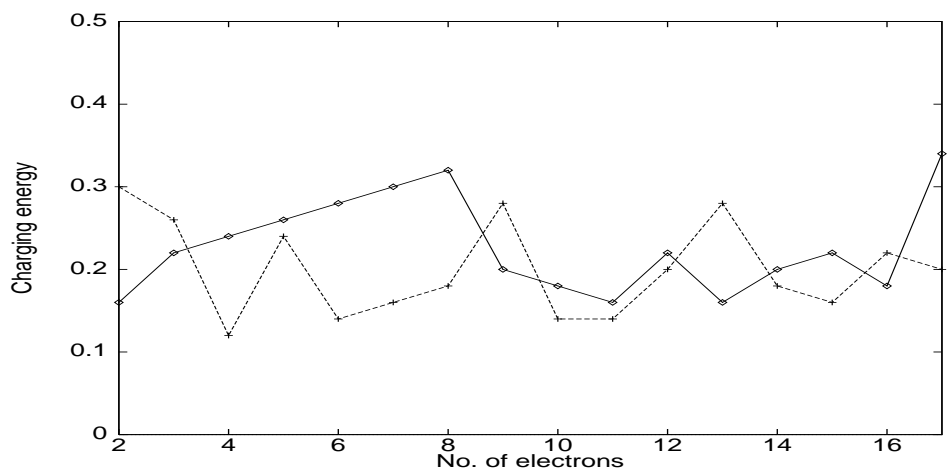
\end{document}